# Virial-potential energy correlation and its relation to the density scaling for quasi-real model systems

K. Koperwas[1,2], A. Grzybowski[1,2], and M. Paluch[1,2]

[1.] University of Silesia in Katowice, Institute of Physics, 75 Pułku Piechoty 1, 41-500 Chorzów, Poland

[2.] Silesian Center for Education and Interdisciplinary Research SMCEBI, 75 Pułku Piechoty 1a, 41-500 Chorzów, Poland

* corresponding author: kajetan.koperwas@us.edu.pl

## ABSTRACT

In this letter, we examine the virial and the potential energy correlation for the quasi-real model system. This correlation constitutes the framework of the theory of the isomorph in the liquid phase diagram commonly examined using simple-liquids. Interestingly, our results show that for the systems characterized by structural anisotropy and flexible bonds, the instantaneous values of total viral and total potential energy are entirely uncorrelated. It is due to the presence of the intramolecular interactions because the contributions to the virial and potential energy resulting from the intermolecular interactions still exhibit strong linear dependence. Interestingly, in contrast to the results reported for simple-liquids, the slope of the mentioned linear dependence is different than the values of the density scaling exponent. However, our findings show that for the quasi-real materials, the slope of dependence between the virial and potential energy (resulting from the intermolecular interactions) strongly depends on the range of intermolecular distances that are taken into account. Consequently, the value of the slope of the discussed relationship, which enables satisfactory density scaling, can be obtained. Furthermore, we show that the above crucial range of intermolecular distances does not depend on the structure of the system as well as on the thermodynamic conditions.

# I.   ARTICLE

The first report on the density scaling of the real material [1,2], which had been published at the turn of the century, significantly stimulated studies on this property of supercooled liquids. The main reason for continuous researchers' fascination with discussed phenomenon is the fact that it straightforwardly links the thermodynamics and dynamics of liquid basing on the simple relationship,

$$X = \mathcal{F}(Tv^{\gamma}),$$  Eq.(1)

where $X$ is a dynamic quantity characterizing the system (e.g., structural relaxation time, viscosity or diffusion constant), $T$ is a temperature, $v$ denotes specific volume and $\gamma$ is material dependent constant. The extensive experimental works confirm that the presented form of the scaling is successfully fulfilled for more than 100 materials. [3] It must also be noted that despite remarkable universality, the form of scaling has another great virtue. The density scaling gives insight into the nature of intermolecular interactions occurring within the system because the scaling exponent, $\gamma$, is directly related to the repulsive part of intermolecular potential. [4–8] The latter implies that the density scaling is reflected in liquid properties such as reported virial and potential-energy correlation [9,10], pressure densification [11] as well as in the physical aging of the glasses. [12]

Among the aforementioned features of supercooled liquids, the virial and potential-energy correlation deserves for particular attention because it constitutes the framework of the isomorphs concept and relating to it the R-simple (Roskilde-simple) liquids. [13] Moreover, it finally led to the proposition of the redefinition of the classical term of the simple-liquids. [14] R-simple systems exhibits strong correlation between fluctuations of virial and potential-energy, where the $\gamma$ is a proportionality constant. [10,15–18] These liquids possess curves in their phase diagram linking isomorphic states at which several dynamic and statistic properties are identical. Consequently, the commonly examine particle distribution functions, normalized time-autocorrelation functions, as well as the transport coefficients, are invariant along the isomorphs, when they are expressed in so-called reduced units. Since the difference between scaling employing unreduced and reduced units is negligible in the supercooled regime, the density scaling rule, given by Eq. (1), is commonly fulfilled for the model as well as real liquids. Therefore, it is not surprising that the huge scientific effort has been done to examine the virial $W$ and potential energy $U$ correlation. The computational experiments on model systems revealed that in general $\gamma$ varies with the thermodynamic conditions [6,19]. However, given that a lot of real liquids accurately fulfill density scaling rule with a constant value of $\gamma$ even for a very wide temperature-pressure range [20,21], the passionate debate on the constancy of $\gamma$ is permanently conducting in the literature. [8,22–25]

Although the exact definition of R-simple liquids is introduced in Ref. [13] their initial name, i.e., 'strongly correlating liquids', is used in the series of five papers devoted to the pressure-energy correlations in liquids. [10,15–18] In the first paper, the Authors introduced the strongly correlated liquids as those exhibiting the strong correlation between fluctuations of virial $\Delta W$ and potential-energy $\Delta U$. The correlation is suggested to be quantified by the Pearson coefficient $R$ of the equilibrium

$$R = \frac{\langle \Delta W \Delta U \rangle}{\sqrt{\langle (\Delta W)^2 \rangle \langle (\Delta U)^2 \rangle}},$$ 

Eq. (2)

where and $\Delta$ denotes instantaneous value of given quantity minus its average value and $\langle \ \ \rangle$ means constant-volume canonical averages. [10] Liquids which exhibit $R \geq 0.9$ are recognized as strongly corelated and for them $\Delta W$ is a linear function of $\Delta U$ with the slope equal to $\gamma$. The latter can be directly calculated from [17]

$$\gamma = \frac{\langle \Delta W \Delta U \rangle}{\langle (\Delta U)^2 \rangle},$$ 

Eq. (3)

which simultaneously gives the least-squared of linear regression best fit slope of $W(U)$. However, at this point, we would like to recall that the perfect correlation between virial and potential-energy fluctuations is a case for the system with pure inverse power-law (IPL) pair potential [14], for which pair potential is proportional to $r_{ij}^{-n}$ ($r_{ij}$ is a distance between two molecules, $n$ is a potential parameter). Mentioned fact results directly from internal virial definition,

$$W = -\frac{1}{3}\sum_{i=1}^{N} \boldsymbol{r}_i \; \boldsymbol{\nabla}_i \boldsymbol{U},$$

Eq.(4)

where $r_i$ is a position of i-*th* particle and $U$ is a total potential energy. [26,27] Nevertheless, considering more realistic models, e.g., system described by (standard for liquid) Lennard-Jones potential, one should take into account that the vast majority of $\Delta W$ and $\Delta U$ comes from molecules separated by the relatively short distances, i.e., distances at which Lennard-Jones potential can be accurately approximated by IPL. Hence, systems with an attractive part of intermolecular potential may also exhibit strong correlation. [10] This situation could be referred to the van der Walls liquids or ionic liquids. [3,28] The different scenario might be observed in the case of associated liquids. The presence of the hydrogen interactions essentially modifies intermolecular potential leading to the break of the discussed correlation. [29,30] However, in the case of the real materials one crucial problem must be noted. The direct experimental examination of the correlation between $\Delta W$ and $\Delta U$ is not accessible, and therefore its existence can be concluded only on the base of its consequences, e.g., validation of the density scaling (defined by Eq. (1)).

In this letter, we unify the results of computational studies made on simple-model systems with those obtained by examinations of the real materials. Basing on computer simulations of quasi-real molecules, which exhibits the simplicity of the common model systems but simultaneously mimics the crucial features of the real molecules, we detailed examine the correlation between instantaneous $W$ and $U$. Our findings show that, in general, $\Delta W$ and $\Delta U$ are entirely uncorrelated if one considers "realistic" molecules. However, instantaneous contributions to $W$ and $U$ originating from intermolecular interactions might still exhibit strong mutual dependence. Interestingly, the value of the slope of this linear relationship is different than the density scaling exponent. Hence, the presented herein observations not only change the general understanding of $WU$ correlation, but they also question its direct relation to the density scaling.

The existence of isomorphs naturally is connected with the form of the intermolecular potential of simple-liquids. [26,31–33] Nevertheless, the reasons for the existence of a strong $WU$ correlation for real materials are not evident. It is mainly because the real molecules possess anisotropic shape, which makes that IPL cannot describe their (anisotropic) intermolecular potential. Taking this fact into account we would like to briefly recall that the correlation between $\Delta W$ and $\Delta U$ has been previously examined for a few model systems comprised of molecules of non-spherical shape. Performed research revealed that the asymmetric and symmetric dumbbell shaped molecules [34,35], Lewis and Wahnström model of ortho-terphenyl (OTP) [36,37], and freely joined chain of atoms [11,38], exhibit the strong $\Delta W$ and $\Delta U$ correlation and they obey the density scaling law. [5,11,38] However, at this point, we have to note that all of those systems possess rigid bonds, which is crucial for $WU$ correlation because bond interactions contribute to the virial as well as to the potential energy of the system. The problem is taken in Ref. [35], where contribution to $W$ resulted from the constraint of bonds is estimated for rigid dumbbells and entirely rigid model of OTP. The exclusion of contribution resulted from constraints decrease $\gamma$ values and increases $R$ value. Hence, it improves $\Delta W$ and $\Delta U$ correlation. However, the real molecules typically possess many flexible bonds, as well as angles and planes, which energies also contribute $\Delta U$.

Fortunately, the given impact to $W$ and $U$ can be directly calculated in computational experiments. In this way we use our recently proposed model of quasi-real system, i.e., rhombus like molecules system (RLMS). [7,8,39] The interactions between non-bonded and bonded atoms are set using the parameters of the OPLSAA force field defined for carbon atoms from aromatic ring. [40] Basing on our previous results [7,8] we chose 3 isochoric conditions, i.e., conditions at which molecular volume $v_m = V/N$ ($N$ is a number of molecules, $V$ is a volume of the system) equals 0.075, 0.085 and 0.095nm³. Subsequently, we simulated RLMS at conditions of constant temperature and volume using Nose-Hover thermostat implemented in GROMACS software. [41–46] Data are collected through the half of a total simulation time, which is 10ns (time step equals 0.001ps). The applied cut-off for intermolecular interactions is set to distance $r_c$ =1.065nm, which is 3 times longer than $\sigma$ parameter of LJ potential describing non-bonded interactions. The chosen temperatures vary from 50 to 200K, which results in a range of pressures equals to 1.2GPa (the temperature dependence of the pressure is shown in Supplemental Material).

In order to confirm the density scaling for RLMS, the diffusion constants, $D$ (determined from mean-square displacement), and relaxation times, $\tau$ (estimated on the base of incoherent intermediate scattering function of molecules centers of mass) expressed in the reduce units (, which are denoted by * and defined as $D^* = \left(v_m^{-1/3}\sqrt{m/k_B T}\right)D$ and $\tau^* = \tau/\left(\sqrt{m/k_B T}v_m^{1/3}\right)$, where $m$ is molecule mass and $k_B$ is Boltzmann constant) are plotted as a function of $Tv_m^{\gamma}$ in Fig. 1. As one can see, accordingly, to the isomorph theory [17] $D^*$ and $\tau^*$ accurately scales with the same $\gamma = 6.173$. The value of the density scaling exponent had been estimated using resulted from Eq. (1) the linear dependence of $log_{10}(T)$ on $log_{10}(v_m)$ at constant value of $D^*$, see Ref. [7,8] for details.

Since the RLMS satisfies the density scaling and the value of the density scaling is known, we can examine the correlation between instantaneous $W$ and $U$. However, as we already mentioned the total potential energy $U_{total}$ of RLMS consists of the term related to the interaction between non-bonded and bonded atoms. Hence, the potentials of intermolecular, bond, bond-angle, and dihedral-angle interactions must be taken into account. At this point we have to stress that, if scalar $W$ is considered, the contribution of any angle-dependent term is zero. [47] Hence, $U_{total} = U_{LJ} + U_{bond} + U_{angle} + U_{dihedral}$, whereas $W_{total} = W_{LJ} + W_{bond}$.

The $WU$ correlation is examined on the example of following thermodynamic conditions, $T =200\text{K}$ and $v_m =0.0075\text{nm}^3$. Interestingly, in the panel Fig. 2a one can clearly see that there is no any correlation between instantaneous $U_{total}$ and $W_{total}$. On the other hand, instantaneous values of $W_{LJ}$ and $U_{LJ}$ resulted from non-bonded interactions are almost perfectly correlated, R=0.993, see Fig. 2b. Hence, we can suspect that $W_{LJ}U_{LJ}$ correlation is broken by the contributions originating from intramolecular interactions. It is due to that the intramolecular interactions are described by harmonic potentials, which cannot be approximated by IPL. It has to be noted that harmonic form of intramolecular potentials is not only matter of choice but it possesses theoretical frameworks. Consequently, none correlation can be expected between instantaneous contributions to $W$ and $U$ from intramolecular interactions, see Fig. 2c where results for bond interactions are presented. Moreover, it can be seen in Fig. 2d that addition of even one intramolecular interaction to $W$ and $U$ completely destructs correlation between them. Summarizing, from Fig. 2 it is evident that $W_{total}U_{total}$ correlation cannot hold for real liquid. Consequently, only intermolecular interaction would be responsible for the density scaling, if its relation with $WU$ correlation is valid for the real materials. This conclusion is especially intriguing if one realizes that the density scaling persists despite that the absolute value of the ratio between intra- and inter-molecular potential energies is about 0.5, i.e., $\left|(U_{total} - U_{LJ})/U_{LJ}\right| \approx 0.5$.

In Fig. 3a we present $R$ and $\gamma_{LJ}$ values estimated on the base of $W_{LJ}U_{LJ}$ correlation for all studied thermodynamic conditions. It is worth noting that only at $T =60\text{K}$ and $v_m =0.0095\text{nm}^3$, $R < 0.9$ and hence it does not satisfy the proposed definition of R-simple liquids. Nevertheless, as we present in Fig. 1, $\tau^*$ and $D^*$ determined at mentioned thermodynamic conditions can be accurately scaled together with all remaining data. However, the most puzzling observation which can be drawn from Fig. 3a is that although $V_{LJ}$ and $U_{LJ}$ are almost perfectly correlated, none of $\gamma_{LJ}$ values calculated from Eq. (3) exceeds 6.0. It implies that the average $\gamma_{LJ}$ cannot be equal (or even close) to expected 6.173. Consequently, density scaling using values shown in Fig. 3a is not valid, see Supplemental Material where density scaling with state dependent $\gamma_{LJ}$, average for each isochrone and average for all studied thermodynamic conditions are presented. The key finding is that none of those scaling can be recognized as valid. Hence, one can suspect that despite the correlation between $V_{LJ}$ and $U_{LJ}$ occurs for the quasi-real liquids it is not directly responsible for its density scaling. On the other hand, the all progress which has been achieved due to the studies on the simple-liquids leading to constitution that $W = \gamma U + const.$ cannot be simply ignored. Therefore, we would like to again pointed out the cardinal difference between typical simple-liquid and the real material, which is the shape and then the interaction anisotropy. Since simple-liquid are comprised of one-atomic molecules their intermolecular potential is entirely spherically symmetrical. Then it does not vary when mutual positions of molecules changes. It makes that it can be approximated by an inverse power law with constant parameters independently on the mutual orientations of molecules and distance between them. As a consequence, atoms occupy positions determined solely by atom-atom intermolecular potential. This situation does not take place for real materials and for considered herein quasi-reals system. In these cases, interactions between molecules result from many atom-atom interactions. It means that intermolecular potential depends on mutual orientations of molecules. As a consequence, at different intermolecular distances different form of the most energetically optimal potential may be expected. Taking the above into account we test the dependence of $\gamma_{LJ}$ values on the interatomic distance. In this purpose, we determine $\gamma_{LJ}$ considering only interactions occuring within the sphere of radius $r_{sphere}$ starting from a given atom. Importantly, described analysis implies that the number of atoms giving impact to the $V_{LJ}U_{LJ}$ correlation is not identical for each $r_{sphere}$. Therefore, we divided the obtained $V_{LJ}$ and $U_{LJ}$ by a number of considered interactions. The results are presented in Fig. 3b, where notable oscillations of $\gamma_{LJ}$ during increasing of $r_{sphere}$ can be observed. However, again even the highest value of $\gamma_{LJ}$ is considerably smaller than 6.173. Nevertheless, it is worth to mention that $\gamma_{LJ}$ initially increases up to its maximal value (a small minimum can be observed but it does not occur for all studied thermodynamic conditions, result not presented), which suggests that omitting the shortest interatomic distances would lead to higher values of $\gamma_{LJ}$. Consequently, we decided to consider subsequent ranges of distances characterized by 0.02nm of width. Then $\gamma_{LJ}$ can be expressed as a function of the position of the range center, $r_{rc}$, see Fig. 3c. Consistently, with the results shown in Fig. 3b, at short intermolecular distances the increase in interatomic distance causes a gain in $\gamma_{LJ}$ values. It is worth mentioning that for consecutive ranges the obtained values are even a few times higher than 6.173 (result not presented). Interestingly, for all boundary thermodynamic conditions, estimated dependences are almost identical. It implies two critical consequences. First, the effective range for $\gamma_{LJ}$ estimation is independent on the thermodynamic conditions. Second, taking into account that the radial distribution functions (RDF) considerably differ at analyzed boundary thermodynamic conditions (see Supplemental Material) it is also not related to the structure of the system. Thus, the positions of ranges centers and their widths seem to be inherent part of the molecular structure.

The centers of the ranges corresponding to $\gamma_{LJ}$=6.173 can be estimated describing $\gamma_{LJ}(r_{rc})$ by the following function, $\gamma_{LJ} = A \cdot exp(r_{rc}/B) + C$, where $A, B, C$ are the fit parameters. The solid lines in Fig. 3c represents results. As one can see, the predicted centers of the sought-after range are obtained at practically the same $r_{rc}$. Estimated $r_{rc}$ values vary from 0.3572nm ($T$ =50K and $v_m$ =0.0085nm³) to 0.3578nm ($T$ =200K and $v_m$ =0.0075nm³) and correspond to the interatomic distances shorter or comparable to the position of the first peak of RDF for atoms (see Supplemental Material). Thus, the interactions taking place between very closest atoms seems to be responsible for the density scaling.

Subsequently, consistently with our previous test, we established $\gamma_{rc}$ taking into account interatomic distances within ranges characterized by 0.02nm of width, and center in predicted $r_{rc}$. The smallest $\gamma_{rc}$=6.119$\pm$0.009 and it established for $T$ =50K and $v_m$ =0.0085nm³, whereas the highest one is registered for $T$ =200K and $v_m$ =0.0075nm³ and equals 6.160$\pm$0.009. Hence, all values are very close to each other. Additionally, it is worth to mention that for all boundary thermodynamic conditions $R$>0.999. Thus, within considered ranges instantaneous $V_{LJ}$ and $U_{LJ}$ are almost perfectly correlated.

Since all obtained $\gamma_{rc}$ are close to 6.173 and their variation is much smaller than the values presented in Fig. 3a, one can expected that $\gamma_{rc}$ will lead to satisfactory density scaling of $\tau^*$ and $D^*$. Due to the fact that all obtained $\gamma_{rc}$ are almost identical, we recognize that estimation of $\gamma_{rc}$ for all thermodynamic conditions is not necessary. Instead, basing on the result for boundary thermodynamic we calculate the mean value of $\gamma_{rc}$=6.138. As it is presented in Fig. 4, performed density scaling is highly accurate for both $\tau^*$ and $D^*$. Concluding, the analysis of the $W_{LJ}U_{LJ}$ correlation could still be useful method for establish the density scaling exponent value. However, one has to remember that a specific range of interatomic distances must be considered. Then, instead of performing complex studies, which lead to the determination of some effective intermolecular potential and subsequent approximate its part by the IPL (, which is analogic to method valid for simple-liquid), one can directly consider the dependence of instantaneous values of $W_{LJ}$ on $U_{LJ}$ at only one thermodynamic condition.

Summarizing, basing on the results for RLMS we have shown that for quasi-real model systems, the $WU$ correlation is not fulfilled. Consequently, a similar situation should be expected for the real liquids instead of that which is commonly suggesting in the literature.

However, contributions to the virial and potential energy resulted from intermolecular interactions ($W_{LJ}$ and $U_{LJ}$) still exhibit the strong mutual dependence. Through the wide range of examined thermodynamic conditions, the correlation coefficient only ones falls below the requested value of 0.9. Unfortunately, in contrast to previously studied simple model systems, the evident linear $W_{LJ}U_{LJ}$ correlation is not characterized by the slope, which leads to satisfactory density scaling. The reason for mentioned difference can be the structural anisotropy of the quasi-real molecules, which makes that the effective intermolecular potential exhibits complex behavior. As a consequence, the effective intermolecular potential cannot be described by single IPL at a wide range of intermolecular distances. Consistently to this hypothesis, we found significant variations of $\gamma_{LJ}$ when different intermolecular distances are considered. Interestingly, $\gamma_{LJ}$=6.173, which enables accurate density scaling, is achieved at almost identical distances independently on the thermodynamic conditions. Even more, we can conclude that the discussed range of distances cannot be determined on the base of structure because the RDFs calculated at studied thermodynamic conditions notably vary between themselves. Hence, the ranges of the intermolecular distances, which are crucial for the density scaling, seem to be an inherent part of the molecular structure. At this point, we would like to note that the proposed herein method should not be treated as a final. The arbitrarily assumed by us the width of ranges can be modified. Then also their centers might be different. Nevertheless, the conclusion on the independence of the positions of the ranges centers on thermodynamic conditions and the structure should still be held. Thus, only when one possesses the knowledge about discussed ranges, the analysis of the $W_{LJ}U_{LJ}$ correlation could be and useful method for the determination of the density scaling exponent. Consequently, this key conclusion put new insight into the nature of the isomorphs for the real materials.

## II.    ACKNOWLEDGEMENTS

The authors are deeply grateful for the financial support by the National Science Centre of Poland within the framework of the Maestro10 project (Grant No. UMO-2018/30/A/ST3/00323).

## III.    REFERENCES

[1]    A. Tölle, H. Schober, J. Wuttke, O. G. Randl, and F. Fujara, Phys. Rev. Lett. **80**, 2374 (1998).

[2]    A. Tölle, Reports Prog. Phys. **64**, 1473 (2001).

[3]    C. M. Roland, S. Hensel-Bielowka, M. Paluch, and R. Casalini, Reports Prog. Phys. **68**, 1405 (2005).

[4]    D. Coslovich and C. M. Roland, J. Phys. Chem. B **112**, 1329 (2008).

[5]    T. B. Schrøder, U. R. Pedersen, N. P. Bailey, S. Toxvaerd, and J. C. Dyre, Phys. Rev. E **80**, 041502 (2009).

[6]    A. Grzybowski, K. Koperwas, and M. Paluch, Phys. Rev. E **86**, 031501 (2012).

[7]    K. Koperwas, A. Grzybowski, and M. Paluch, J. Chem. Phys. **150**, 014501 (2019).

[8]    K. Koperwas, A. Grzybowski, and M. Paluch, Phys. Rev. E **101**, 012613 (2020).

[9]    U. R. Pedersen, N. P. Bailey, T. B. Schrøder, and J. C. Dyre, Phys. Rev. Lett. **100**, 015701 (2008).

[10]    N. P. Bailey, U. R. Pedersen, N. Gnan, T. B. Schrøder, and J. C. Dyre, J. Chem. Phys. **129**, 184507 (2008).

[11]    D. Fragiadakis and C. M. Roland, J. Chem. Phys. **147**, 084508 (2017).

[12]    N. Gnan, C. Maggi, T. B. Schrøder, and J. C. Dyre, Phys. Rev. Lett. **104**, 125902 (2010).

[13]    T. B. Schrøder and J. C. Dyre, J. Chem. Phys. **141**, 204502 (2014).

[14]    T. S. Ingebrigtsen, T. B. Schrøder, and J. C. Dyre, Phys. Rev. X **2**, 011011 (2012).

[15]    N. P. Bailey, U. R. Pedersen, N. Gnan, T. B. Schrøder, and J. C. Dyre, J. Chem. Phys. **129**, 184508 (2008).

[16]    T. B. Schrøder, N. P. Bailey, U. R. Pedersen, N. Gnan, and J. C. Dyre, J. Chem. Phys. **131**, 234503 (2009).

[17]    N. Gnan, T. B. Schrøder, U. R. Pedersen, N. P. Bailey, and J. C. Dyre, J. Chem. Phys. **131**, 234504 (2009).

[18]    T. B. Schrøder, N. Gnan, U. R. Pedersen, N. P. Bailey, and J. C. Dyre, J. Chem. Phys. **134**, (2011).

[19]    U. R. Pedersen, N. Gnan, N. P. Bailey, T. B. Schrøder, and J. C. Dyre, J. Non. Cryst. Solids **357**, 320 (2011).

[20]    E. H. Abramson, J. Phys. Chem. B **118**, 11792 (2014).

[21]    T. C. Ransom and W. F. Oliver, Phys. Rev. Lett. **119**, 025702 (2017).

[22]    A. Sanz, T. Hecksher, H. W. Hansen, J. C. Dyre, K. Niss, and U. R. Pedersen, Phys. Rev. Lett. **122**, 055501 (2019).

[23]    T. C. Ransom, R. Casalini, D. Fragiadakis, A. P. Holt, and C. M. Roland, Phys. Rev. Lett. **123**, 189601 (2019).

[24]    Z. Wojnarowska, M. Musial, M. Dzida, and M. Paluch, Phys. Rev. Lett. **123**, 125702 (2019).

[25]    T. C. Ransom, R. Casalini, D. Fragiadakis, and C. M. Roland, J. Chem. Phys. **151**, 174501 (2019).

[26]   J. P. Hansen and I. R. McDonald, *Theory of Simple Liquids* (Elsevier Academic Press, London; Burlington, 2006).

[27]   M. P. Allen and D. J. Tildesley, *Computer Simulation of Liquids* (Oxford University Press, 2017).

[28]   A. S. Pensado, A. A. H. Pádua, M. J. P. Comuñas, and J. Fernández, J. Phys. Chem. B **112**, 5563 (2008).

[29]   C. M. Roland, S. Bair, and R. Casalini, J. Chem. Phys. **125**, 124508 (2006).

[30]   S. Pawlus, M. Paluch, and A. Grzybowski, J. Chem. Phys. **134**, 041103 (2011).

[31]   B. Bernu, J. P. Hansen, Y. Hiwatari, and G. Pastore, Phys. Rev. A **36**, 4891 (1987).

[32]   J. N. Roux, J. L. Barrat, and J. P. Hansen, J. Phys. Condens. Matter **1**, 7171 (1989).

[33]   J. Barrat and A. Latz, J. Phys. Condens. Matter **2**, 4289 (1990).

[34]   U. R. Pedersen, T. Christensen, T. B. Schrøder, and J. C. Dyre, Phys. Rev. E **77**, 011201 (2008).

[35]   T. S. Ingebrigtsen, T. B. Schrøder, and J. C. Dyre, J. Phys. Chem. B **116**, 1018 (2012).

[36]   L. J. Lewis and G. Wahnström, Solid State Commun. **86**, 295 (1993).

[37]   L. J. Lewis and G. Wahnström, Phys. Rev. E **50**, 3865 (1994).

[38]   A. A. Veldhorst, J. C. Dyre, and T. B. Schrøder, J. Chem. Phys. **141**, 54904 (2014).

[39]   K. Koperwas, K. Adrjanowicz, A. Grzybowski, and M. Paluch, Sci. Rep. **10**, 283 (2020).

[40]   W. L. Jorgensen, D. S. Maxwell, and J. Tirado-Rives, J. Am. Chem. Soc. **118**, 11225 (1996).

[41]   H. J. C. Berendsen, D. van der Spoel, and R. van Drunen, Comput. Phys. Commun. **91**, 43 (1995).

[42]   M. J. Abraham, T. Murtola, R. Schulz, S. Páll, J. C. Smith, B. Hess, and E. Lindahl, SoftwareX **1–2**, 19 (2015).

[43]   S. Páll, M. J. Abraham, C. Kutzner, B. Hess, and E. Lindahl, in *Solving Softw. Challenges Exascale Int. Conf. Exascale Appl. Software, EASC 2014*, edited by S. Markidis and E. Laure (Springer International Publishing, Cham, 2015), pp. 3–27.

[44]   S. Nosé, Mol. Phys. **52**, 255 (1984).

[45]   S. Nosé, J. Chem. Phys. **81**, 511 (1984).

[46]   W. G. Hoover, Phys. Rev. A **31**, 1695 (1985).

[47]   H. Bekker, Molecular Dynamics Simlation Methods Revised, 1996.

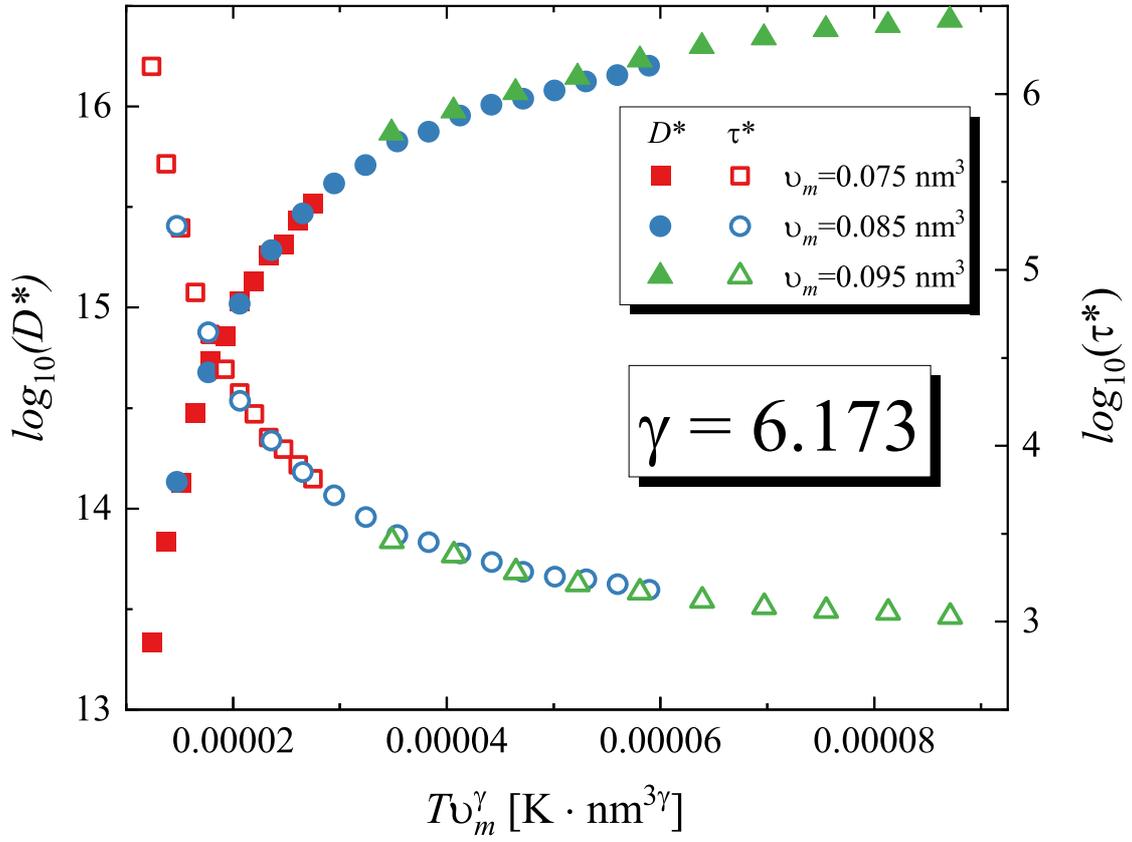

Fig. 1 (color online)

The density scaling for RLMS with constant value of $\gamma$ determined as a slope of linear dependence of $log_{10}(T)$ on $log_{10}(v_m)$ at constant value of $D^*$

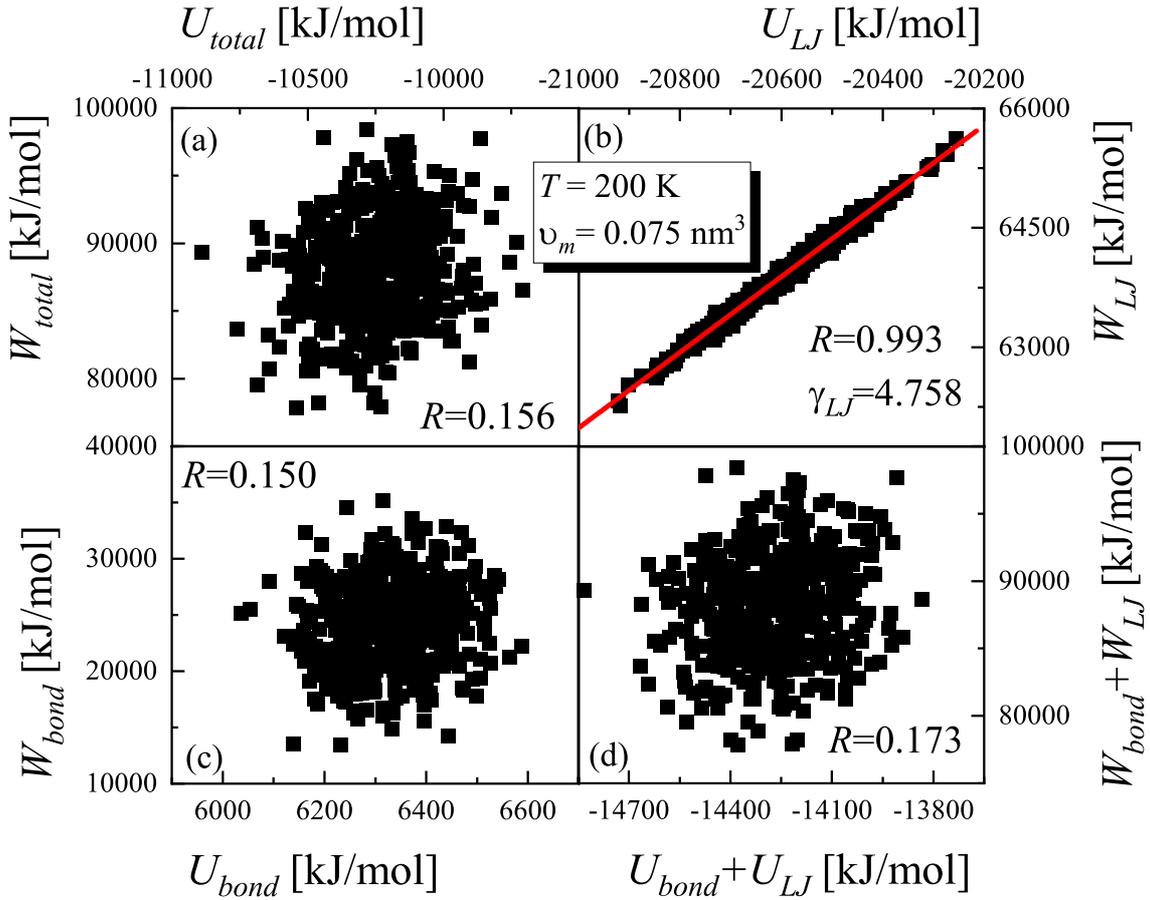

Fig. 2 (color online)

The dependence of the instantaneous values of total virial on total potential energy is presented in panel (a). The contributions to $W_{total}$ and $U_{total}$ resulted from intermolecular interactions and from bond interactions are shown in panel (b) and (c) respectively. The red line in panel (b) represents the fit linear function. In the panel (d) the sums of contributions shown in panels (b) and (c) are depicted.

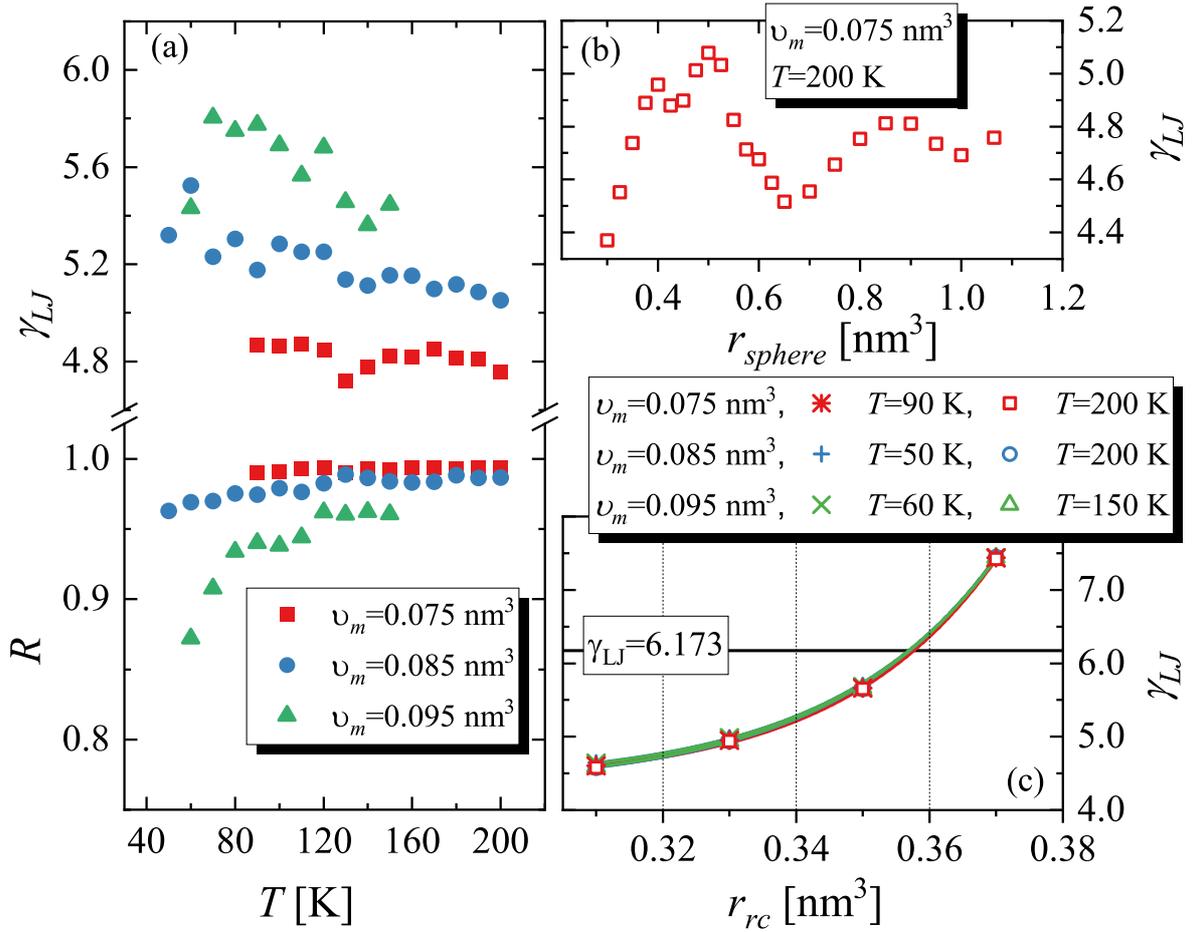

Fig. 3 (color online)

The calculated at various thermodynamic conditions values of $\gamma_{LJ}$ (top) and corresponding to them R (bottom) are shown in the panel (a). In the panel (b) the dependence of $\gamma_{LJ}$ on the radius of the sphere embracing interacted atoms is presented. In panel (c) the values of the $\gamma_{LJ}$ obtained for the interatomic distance located within the ranges of width equals 0.02nm is plotted as a function of the positions of the ranges centers. The vertical dotted lines represent the boarders between subsequent ranges. Solid lines are fits to the exponential function of the obtained results.

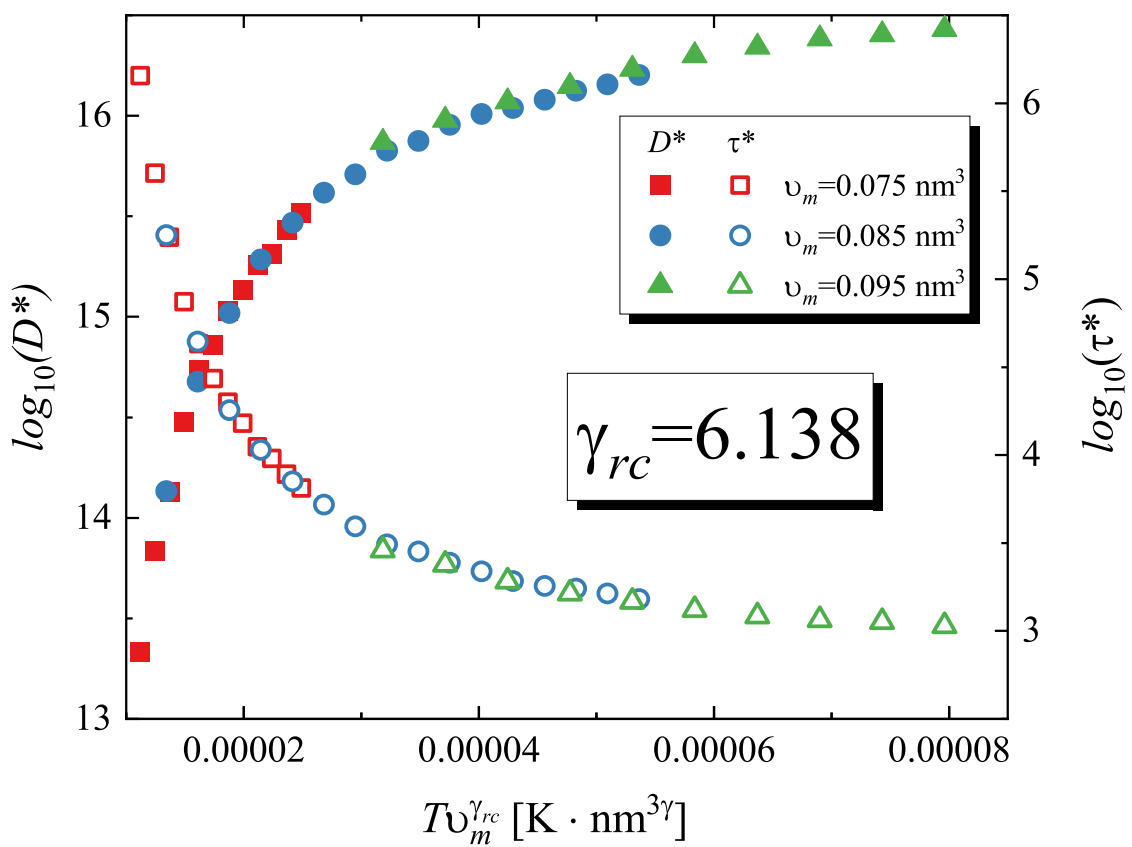

Fig. 4 (color online)

The density scaling for RLMS with a constant value of $\gamma_{rc}$ determined from the analysis of the correlation between instantaneous $V_{LJ}$ and $U_{LJ}$ considering exclusively atoms separated by the specific distances.